\documentclass[onecolumn]{IEEEtran}

  \usepackage[pdftex]{graphicx}

  \DeclareGraphicsExtensions{.eps,.pdf,.jpg,.png}

\usepackage{amsmath}
\usepackage{amssymb}
\usepackage{amsthm}
\usepackage{fancyhdr}
\usepackage[update,prepend]{epstopdf}
\usepackage{paralist}
\usepackage{subfig}

\usepackage{array}
\usepackage{verbatim}
\usepackage{subfig}
\usepackage[update,prepend]{epstopdf}
\usepackage{multirow}

\newtheorem{theorem}{Theorem}
\newtheorem{definition}{Definition}

\begin{document}
\title{Mathematical Foundations for Information Theory in Diffusion-Based Molecular Communication}
\author{Ya-Ping Hsieh and Ping-Cheng Yeh,~\IEEEmembership{Member,~IEEE,}
\thanks{Ya-Ping Hsieh and Ping-Cheng Yeh are with the Graduate Institute Communication, National Taiwan University, Taipei, Taiwan, e-mail: \{b95901020@ntu.edu.tw\}, \{pcyeh@cc.ee.ntu.edu.tw\}.}}



\maketitle

\begin{abstract}
Molecular communication emerges as a promising communication paradigm for nanotechnology. However, solid mathematical foundations for information-theoretic analysis of molecular communication have not yet been built. In particular, no one has ever proven that the channel coding theorem applies to molecular communication, and no relationship between information rate capacity (maximum mutual information) and code rate capacity (supremum achievable code rate) has been established. In this paper, we focus on a major sub-class of molecular communication -- the diffusion-based molecular communication. We provide solid mathematical foundations for information theory in diffusion-based molecular communication by creating a general diffusion-based molecular channel model in measure-theoretic form and prove its channel coding theorems. Various equivalence relationships between statistical and operational definitions of channel capacity are also established, including the most classic information rate capacity and code rate capacity. As byproducts, we have shown that the diffusion-based molecular channel is with ``asymptotically decreasing input memory and anticipation'' and ``$\bar{d}$-continuous''. Other properties of diffusion-based molecular channel such as stationarity or ergodicity are also proven.
\end{abstract}


\IEEEpeerreviewmaketitle

\section{Introduction}
\IEEEPARstart{M}{olecular} communication is a recently developed communication paradigm whose communication process involves ``transmission and reception of information encoded in molecules'' \cite{akyildiznanonetworks}. It is born in the study of nano-scale communication in nanonetworks \cite{akyildiznanonetworks} where the applicability of classical electromagnetic communication is limited by several constraints and a novel solution is called for. Inspired by the biological communication process such as the intra- and inter-cellular communication by cells \cite{citeulike:358722} and the pheromone diffusion by insects \cite{Bossert1963443}, the researchers have created various message-carrying molecules and their corresponding receptors. Since the molecules are themselves at the nano-scale, the difficulty of nano-scale communication is solved by building up communication media between the transmitter and the receptor. Moreover, since the major potential application of nanonetworks is implemented in the organisms body (e.g., organ monitoring in human body), the compatibility of the bioinspired communication process makes it a perfect choice to consider the encoding, transmission, and decoding of messages in molecules. Excited by the reasons above (and beyond), the researchers have treated the molecular communication as one of the mainstream subjects of nanotechnology nowadays.

Among the various proposed system blue maps for molecular communication, a specific class of system called the ``diffusion-based molecular system'' draws lots of interest due to the universality of {\em mass transport phenomena} in molecular communication. The mass transport phenomena refer to the propagation of encoded molecules in the communication media. In most cases, the communication media of molecular communication are modeled as fluid. As is well-known, the motion of molecules in fluid media is governed by the ``diffusion process''. The universality of mass transport phenomena not only make the study of diffusion-based system natural, but at the same time guarantee the widest possible generality of research results. To this end, most efforts in molecular communication are devoted to diffusion-based systems, both in practical system design and information-theoretic analysis point of view. For example, for the practical system design, the physical phenomena of bacteria chemotaxis, pheromone diffusion \cite{ParcerisaGine:2009:MCO}, or calcium diffusion \cite{Nakano05molecularcommunication} are adopted, and there are many existing systems operating on these phenomena such as \cite{6364540}, \cite{Hsieh2012}, and \cite{6364963}. The information-theoretic analysis of diffusion-based molecular system are also conducted in \cite{6034228}, \cite{5935214}, and \cite{6305481}. 

Despite of this much effort for diffusion-based molecular communication, two important things are still lacking. First, {\em an unified channel model for diffusion-based molecular communication is missing.} The existing works on information-theoretic analysis tend to be system-dependent; that is, they are treating different kinds of diffusion-based systems independently and giving their own capacity analysis. So far, the most general result concerning the information rate capacity of diffusion-based molecular communication is probably \cite{6305481} where the authors have cleverly utilized the thermodynamics to derive the information rate capacity in closed form, but the result is only valid for receiver with Poisson noise. Second, and the more important one, {\em no mathematical foundations have been put down for information theory in diffusion-based molecular communication}. In particular, the channel coding theorem is not proven for diffusion-based systems. To state more directly, no one has ever shown that, for the diffusion-based molecular channel, the information rate capacity (maximum mutual information) is equal to the code rate capacity (supremum achievable code rate), where the former is a system statistic of mathematical importance and the latter is the real practical concern. Furthermore, the diffusion-based molecular channels exhibit non-classical randomness (which is explained later in Section IV.A) and {\em no existing channel model apply to diffusion-based molecular channel}. Therefore, the lacking of proof for equivalence between the information rate capacity and code rate capacity renders the purpose of calculating information rate capacity ambiguous, or at least in a shaky situation.

This paper aims at solving the above two problems once and for all. Let us first put molecular communication aside for a moment and recall the history of information theory. Ever since the advent of the paper \cite{shannon1948} by Shannon in 1948, the ``Channel Capacity'' has been the primary subject of every communication systems. In particular, the way that Shannon intervenes the information-theoretic and coding-theoretic quantities for discrete memoryless channels is the one that communication engineers imitate when treating the communication system under consideration. Despite its great novelty and ingenuity, a complete and rigorous mathematical setting for information theory is not laid down until a series of papers by McMillan \cite{McMi53a}, Feinstein \cite{Feinstein1954}, and finally Khinchin \cite{khinchin1957mathematical}, where the definition of ``finite memory channel without anticipation'' is made precise and for such channels the equivalence between the information rate capacity and code rate capacity is established. A key step of \cite{khinchin1957mathematical} is the {\em measure-theoretic formulation for sources and channels}. Khinchin noticed that except for some simple classes of channel (e.g., i.i.d., Markov, ...), the mathematical formulations of sources and channels are most conveniently made in the measure-theoretic form to provide rigorousness and applicability. Later, works in \cite{khinchin1957mathematical} are generalized by many others (e.g., \cite{1054666}, \cite{1056045}, \cite{genecapa}, ...) to include classes of channels other than finite memory channel without anticipation, all under the same proof structure. It is now a communication engineers' consensus that the measure-theoretic formulations of source and channel have struck a perfect balance between the mathematical rigorousness of information theory and the applicability thereof to practical communication systems. Therefore, for any communication system, once its channel abstraction is made in the measure-theoretic form and the corresponding capacity theorems are derived, then a solid mathematical foundation is laid down and applicability of information theory to the system is guaranteed.

In this paper, we shall follow the path of Khinchin \cite{khinchin1957mathematical} to provide the mathematical foundations for information theory in diffusion-based molecular communication. Fortunately, there are already many building blocks from the works of predecessors \cite{1054666}, \cite{1056045}, and \cite{1056074}. From these works, the two crucial concepts of ``channels with asymptotically decreasing input memory and anticipation'' and ``$\bar{d}$-continuous channels'' will be used heavily in our paper. We will give a channel model for diffusion-based molecular communication in the most rigorous language, namely the measure-theoretic one. The channel model we create captures the essential features of all existing diffusion-based molecular systems and all diffusion-based molecular systems to come, thus allowing our results the maximum possible generality. For our channel, we first prove its $\bar{d}$-continuity and then utilize the $\bar{d}$-continuity to establish various equivalence relationships between the statistical and operational definitions of channel capacity, including the equivalence of information rate capacity and the code rate capacity. In particular, the channel coding theorems for diffusion-based molecular channel is given. A solid foundation is therefore laid down and we can use the usual sense of ``channel capacity'' without worry thereafter.

The rest of paper is organized as follows. We state the essential terminologies and notations in Section II. Section III introduces ``channels with asymptotically decreasing input memory and anticipation'' and ``$\bar{d}$-continuous channels'' as well as their properties for our use. Section IV is the main section in which we give the definition of diffusion-based molecular channel and prove capacity theorems, along with some discussions. Finally, we conclude this paper in Section V.

\section{Preliminaries}
\subsection{Sources}
We begin by setting up terminology and notation for sources. If a set $A$ is finite, the symbol $A^{\infty}_{-\infty}$ will denote the set of all doubly infinite sequences $x = (\cdots, x_{-2}, x_{-1}, x_0, x_1, x_2, \cdots)$, $x_i \in A$ for all $i$. The notation $x^n_m$ denotes the finite sequence $(x_m, x_{m+1}, \cdots, x_n)$ and $A^n_m$ stands for the set of all such sequences. Given a sequence $a^n_m \in A^n_m$, there corresponds a cylinder set (or rectangle set): 
$$
C(a^n_m) = \{ x\in A^{\infty}_{-\infty} | x_i = a_i \text{ for } m \leq i \leq n\}.
$$
The Borel field generated by all cylinder sets in $A^{\infty}_{-\infty}$ will be denoted by $\mathcal{B}_A$. 

A source is a probability measure $\mu$ on the measurable space $(A^{\infty}_{-\infty}, \mathcal{B}_A)$ and $A$ is called the alphabet of the source. The probability measure $\mu$ induces a sequence of random variables $\{X_n | n= -\infty, \cdots, \infty\}$ on the probability space $(A^{\infty}_{-\infty}, \mathcal{B}_A, \mu)$ by the formula $X_n(x) = x_n$ for all $x \in A^{\infty}_{-\infty}$. For integers $m$ and $n$ define $\mathcal{B}_A(m,m+n)$ as the Borel field generated by random variables $(X_m, X_{m+1}, \cdots, X_{m+n})$ and we allow $m=-\infty$ and $n = \infty$. The induced sequence $\{X_n\}_n$ is viewed as the transmitted signal for each time $n$, and $\mathcal{B}_A(m,m+n)$ represents all possible events in time $m$ to $m+n$.

Besides $\{X_n\}_n$, $\mu$ also defines a probability measure $\mu^n_m$ on $A^n_m$ by the formula 
$$
\mu^n_m(a^n_m) = \mu(C(a^n_m)), \ \ \ \\ \  a^n_m \in A^n_m.
$$ 
When $m = 0$, $\mu^n_m$ is abbreviated $\mu^n$. Whenever we need to emphasize the alphabet or the induced random variables, we will also use the notation $[A, \mu]$ or $[A, \mu, X]$ to denote a source. 

The (left) shift operator $T$ on $A^{\infty}_{-\infty}$ is defined by $(Tx)_n = x_{n+1}$, where $(Tx)_n$ is the element at the $n$th position. A source $\mu$ is said to be stationary if 
$$
\mu(F) = \mu(TF) = \mu(T^{-1}F), \ \ \ \ F\in \mathcal{B}_A.
$$ A source is $n$-stationary for some positive integer $n$ if $\mu(T^nF) = \mu(F)$ for all $F \in \mathcal{B}_A$. In other words, $\mu$ is stationary if it is shift-invariant with respect to all events, and it is $n$-stationary if it is $n$-times shift-invariant with respect to all events. Obviously stationarity implies $n$-stationarity. If $\mu$ is $n$-stationary for some $n$, we say that $\mu$ is block stationary.

A source $\mu$ is ergodic if for every $F \in \mathcal{B}_A$ such that $TF = F$, $\mu(F)$ is either $0$ or $1$. An important property of the ergodic sources is the following:
\begin{theorem}
For a given source $[A, \mu]$, denote the characteristic function of the set $F \in \mathcal{B}_A$ by $\chi_F(\cdot)$. Then, we have for all $F \in \mathcal{B}_A$ 
$$
\lim_{n\rightarrow \infty} \frac{1}{n}\sum_{k=0}^{n-1}\chi_F(T^k x) = \mu(F) {\text \ \ \ \ \ \     a.e.,}
$$
if and only if for every $F$ such that $TF = F$, $\mu(F)$ is either $0$ or $1$.
\end{theorem}
\begin{IEEEproof}
The necessity is provided by the well-known ``ergodic theorem'' of Birkhoff and the sufficiency is a direct verification of the definition. For complete proof, see, e.g., \cite{khinchin1957mathematical}.
\end{IEEEproof}
The summation appeared in the theorem represents the ``ensemble mean''.  {\bf Theorem 1} therefore states that the ergodic sources are those whose features can be approximated well by ensemble statistics on observable samples.

\subsection{Channels}
A discrete channel is characterized by an input alphabet $A$, an output alphabet $B$, and a set of probability measures $\{\nu_x\}_x$ on $(B^{\infty}_{-\infty}, \mathcal{B}_B)$. The probability measure $\nu_x$ is determined for each $x \in A^{\infty}_{-\infty}$ such that for each $F \in \mathcal{B}_B$ the mapping $x \rightarrow \nu_x(F)$ is $\mathcal{B}_A$-measurable. Each $\nu_x$ is recognized as the output distribution conditioning on the input $x$. We shall refer $[A, \nu, B]$ to such structure and say that $[A, \nu, B]$ is a channel. The channel is stationary if for every $x\in A^{\infty}_{-\infty}$ and $F \in \mathcal{B}_B$ we have
$$
\nu_x(F) = \nu_{Tx}(TF).
$$

Now, consider the case of connecting a source $[A, \mu]$ to a channel $[A, \nu, B]$. We are naturally led to consider the probability measure $\mu\nu$ on the product space $((A\times B)^{\infty}_{-\infty}, \mathcal{B}_{A\times B})$ characterized by
$$
\mu\nu(F\times G) = \int_F \nu_x(G)d\mu(x), \ \ \ \ F\in \mathcal{B}_A, \ \ G \in \mathcal{B}_B.
$$
The event $F\times G$ is interpreted as ``source emits event $F$ and the channel output receives $G$''. If we are interested in the output process, we shall consider the probability measure $\overline{\mu\nu}$ on the output space $(B^{\infty}_{-\infty}, \mathcal{B}_B$):
$$
\overline{\mu\nu}(G) = \mu\nu(A^{\infty}_{-\infty}\times G), \ \ \  G\in \mathcal{B}_B.
$$
The sequence of random variables induced by $\overline{\mu\nu}$ on $(B^{\infty}_{-\infty}, \mathcal{B}_B$) will be denoted by $\{Y_n\}_n$. Elementary analysis shows that if $[A, \mu]$ and $[A, \nu, B]$ are both stationary, then so is $\mu\nu$ and $\overline{\mu\nu}$.

In this paper we will encounter the ``cascade'' of channels. Suppose that we are given two channels $[A, \sigma, Q]$ and $[Q, \eta, B]$. If we have these two channels in cascade, the overall channel $[A, \nu, B]$ is described by
$$
\nu_x(F) = \int_{Q_{-\infty}^{\infty}} \eta_q(F) d\sigma_x(q).
$$

Another important notion of channel is the ergodicity. The channel $\nu$ is ergodic if for any input sequence $x$, $\nu_x$ is ergodic. The following is an useful criterion for examining the ergodicity of a channel:
\begin{theorem}[Strongly Mixing and Ergodicity]
A stationary channel $[A, \nu, B]$ is said to be strongly mixing if for any input sequence $x$ and any two cylinders $F_1, F_2 \in \mathcal{B}_B$, we have
$$
\lim_{n\rightarrow \infty}|\nu_x(T^kF_1 \cap F_2) - \nu_x(T^kF_1)\nu_x(F_2)| \rightarrow 0
$$ where $T$ is the time shift operator. If a channel is strongly mixing, then it is ergodic.
\end{theorem}
\begin{IEEEproof}
See \cite{1054666}.
\end{IEEEproof} Intuitively speaking, a channel is strongly mixing if for any two events separated by a long time at the output, the two events are ``almost'' independent, regardless of the input.

\subsection{Information-Theoretic and Coding-Theoretic Quantities}
We have seen that the source $\mu$ can be viewed as a sequence of random variables $\{X_n\}_n$. Similarly, we can view $\overline{\mu\nu}$ as $\{Y_n\}_n$ and $\mu\nu$ as $\{X_n, Y_n\}_n$. It is well-known that if the processes represented by these sequences of random variables are stationary then their entropy rates exist. Since we shall only consider stationary ergodic sources and stationary channels in this paper, we are free to write
$$
H(\mu) = H(X) = \lim_{N\rightarrow \infty} \frac{H(X_1, X_2, ..., X_N)}{N}
$$
$$
H(\overline{\mu\nu}) = H(Y) = \lim_{N\rightarrow \infty} \frac{H(Y_1, Y_2, ..., Y_N)}{N}
$$
\begin{align*}
H(\mu\nu) &= H(X, Y) \\ &= \lim_{N\rightarrow \infty} \frac{H((X_1, Y_1), (X_2, Y_2), ..., (X_N, Y_N))}{N}.
\end{align*}
The mutual information rate between the input and output is defined as
$$
I(\mu\nu) = H(\mu) + H(\overline{\mu\nu}) - H(\mu\nu)
$$
or, equivalently, 
$$
I(X; Y) = H(X)+H(Y)-H(X,Y).
$$
In the rest of this paper we shall refer to these quantities interchangably in their measure forms and random variable forms.\\

There are many defintions of channel capacity. In the followings we state and itemize them for ease of reference. See \cite{1056045}, \cite{genecapa}, and \cite{1056074} for detail descriptions of these channel capacities.

Given a channel $[A, \nu, B]$, there are several choices of source which we can extremizing the mutual information over. The most common ones are the follows: 

\begin{itemize}
\item The ergodic capacity 
$$
C_e = \sup_{\text{stationary and ergodic $[A, \mu]$}} I(\mu\nu),
$$
\end{itemize}
\begin{itemize}
\item The stationary capacity
$$
C_s = \sup_{\text{stationary $[A, \mu]$}} I(\mu\nu),
$$\end{itemize}
and \begin{itemize}
\item The block stationary capacity
$$
C = \sup_n \ \sup_{n-\text{stationary} [A,\mu]} I(\mu\nu).
$$\end{itemize}

We now state coding-theoretic terminologies. A blocklength $n$ channel block code $\mathcal{E}$ for a noisy channel $[A, \nu, B]$ is a collection of $|\mathcal{E}| = M$ distinct codewords $w_i \in A^n$ and $M$ disjoint decoding sets $\Gamma_i \in \mathcal{B}_B(0,n)$ for each $i$. The rate $R$ of a code is $R = \frac{1}{n}\log M$. A code $\mathcal{E} = \{w_i, \Gamma_i | i=1, 2, \cdots, M\}$ has probability error $\lambda$ if 
$$
\max_{1\leq i \leq M} \sup_{x\in c(w_i)} \nu_x^n(\Gamma^c_i) \leq \lambda
$$
where $\Gamma^c_i$ is the complement of $\Gamma_i$. Such a channel block code is called an $(M, n, \lambda)$ channel code.

We say $R$ is a permissible transmission rate if there exist $(<2^{nR}>, n, \epsilon_n)$ channel codes such that $\epsilon_n \rightarrow 0$ as $n\rightarrow \infty$. The operational channel block coding capacity $C_{cb}$ is defined to be the supreme of all permissible rate:
\begin{itemize} \item The operational channel block coding capacity $$C_{cb} = \sup_{R \text{\ permissible}} R.$$ \end{itemize}

An alternate operational definition of capacity, the operational source/channel block coding capacity $C_{scb}$, is defined as follows. We consider the transmission of an ergodic source $[G, \tau, U]$ over the channel $[A, \nu, B]$. Define a blocklength $n$ source/channel block code as a pair of mappings $\gamma_n : G^n \rightarrow A^n$ and $\psi_n : B^n\rightarrow G^n$. The $\gamma_n$ is called the encoder and $\psi_n$ the decoder. The block encoder $\gamma_n$ implements a mapping on infinite sequences $\gamma : G^\infty \rightarrow A^\infty$ defined by $\gamma(u) = (\cdots, \gamma_n(u_{-n}, \cdots, u_{-1}), \gamma_n(u_0, \cdots, u_{n-1}), \cdots)$. All such encoded sequences can be viewed as another channel input process $[A, \tau\gamma^{-1}]$ characterized by $\tau\gamma^{-1}(F) = \tau(\gamma^{-1}(F))$ for all $F \in \mathcal{B}_A$. The block error probability for the code is 
$$
P_e = Pr(\psi_n(Y^n) \neq U^n)
$$ where $Y^n$ is the random variable induced at the channel output, or, presented in the measure from, 
$$
P_c = 1-P_e = \int \nu^n_{\gamma(u)}(\psi_n^{-1}(U^n(u)))d\tau(u).
$$ A source/channel code of blocklength $n$ and of probability error $P_e$ is called an $(n, P_e)$ source/channel code.

A source $[G, \tau]$ is said to be admissible if there exist $(n, P_{e, n})$ source/channel codes for the source such that $P_{e,n} \rightarrow 0$ as $n \rightarrow \infty$. The operational source/channel block coding capacity $C_{scb}$ is defined as the supreme of the entropy over all admissible stationary ergodic sources:
\begin{itemize} \item The operational source/channel block coding capacity$$
C_{scb} = \sup_{\text{admissible\ } [G, \tau]} H(\tau).
$$\end{itemize}

The last operational capacity definition concerns another function of samples defined over input/output alphabets. The $n$th order sample mutual information $i_n(x^n, y^n)$ is defined by
$$
i_n(x^n, y^n) = \frac{1}{n}\log \frac{\mu\nu^n(x^n, y^n)}{\mu^n(x^n)\overline{\mu\nu}^n(y^n)}.
$$
If $\mu\nu$ is stationary then there exist a measurable shift-invariant nonnegative bounded function $i(x,y)$ such that 
$$
\lim_{n\rightarrow \infty} i_n(X^n(x), Y^n(y)) = i(x,y)
$$ with convergence in $L_1(\mu\nu)$ and
$$
E_{\mu\nu} i = \int i(x,y) d\mu\nu(x,y) = I(\mu\nu).
$$ If, furthermore, $\mu\nu$ is ergodic, then the Shannon-McMillan Theorem states that $i(x, y) = I(\mu\nu)$ $\mu\nu$-almost everywhere. Therefore, for stationary ergodic sources and channels the $n$th order sample mutual information $i_n(x^n, y^n)$ is a good approximation of the ``true'' mutual information $I(\mu\nu)$ provided $n$ is large enough. Define for a stationary channel $\nu$ and for $\lambda \in (0, 1)$ the quantile
$$
C^*(\lambda) = \sup_\mu \ \sup \{r \ | \ \mu\nu(i \leq r) < \lambda \}
$$ where $\mu\nu(i \leq r) = \mu\nu(\{(x, y)\ |\ i(x,y) \leq r\})$ and the outer supreme is taken over all block stationary input process $[A, \mu]$. Our last operational capacity definition, the information quantile capacity $C^*$, is defined by
\begin{itemize}
\item The information quantile capacity
$$
C^* = \lim_{\lambda \rightarrow 0} C^*(\lambda).
$$\end{itemize}

The relation between all these capacities (namely, $C_e, C_s, C, C_{cb}, C_{scb}$, and $C^*$) will be given in Section III, where we need the notion of $d$-continuity to unify these concepts.

\section{The ADIMA Channel and the $\bar{d}$-Distance}
\subsection{ADIMA Channel}
A channel $\nu$ is said to have asymptotically decreasing input memory if given $\epsilon > 0$ there is an integer $m(\epsilon)$ such that, for any $n$ and any output event $F \in \mathcal{B}_B(n, \infty)$,
$$
|\nu_x(F) - \nu_{x'}(F)| \leq \epsilon
$$whenever $x_i = x'_i$ for $i\geq n-m$. A channel $\nu$ is said to have asymptotically decreasing input anticipation if given $\epsilon > 0$ there is an integer $a(\epsilon)$ such that for any $n$ and any output event $F \in \mathcal{B}_B(-\infty, n)$,
$$
|\nu_x(F) - \nu_{x'}(F)| \leq \epsilon
$$ whenever $x_i = x'_i$ for $i \leq n+a$. If a channel has both asymptotically decreasing input memory and input anticipation then we call such channels the ADIMA (asymptotically decreasing input memory and anticipation) channels or channels with ADIMA. Therefore, a channel $\nu$ with ADIMA is one for wich given $\epsilon >0$, there exist integers $m(\epsilon)$ and $a(\epsilon)$ such that for all $k, n$ and all $F \in \mathcal{B}_B(k, k+n)$ we have
$$
|\nu_x(F) - \nu_{x'}(F)| \leq \epsilon
$$ whenever $x_i = x'_i$, $k-m \leq i \leq k+n+a$.

Intuitively speaking, a channel has asymptotically decreasing input memory if for any two input sequences $x$ and $x'$, the conditional channel output distributions $\nu_x$ and $\nu_{x'}$ in the interval $(y_n, y_{n+1}, \cdots)$ are ``close'' provided the two input sequences coincide on the interval $(x_{n-m}, x_{n-m+1}, \cdots)$. Similar interpretation works for channel with asymptotically decreasing input anticipation. The measure of closeness can be made precise in the language of the {\em variational distance}. Given two probability measures $\nu_x^n$ and $\nu_{x'}^n$ on $(B^n, \mathcal{B}_B(0,n))$, the variational distance between the measures is defined by 
$$
v_n(\nu_x^n, \nu_{x'}^n)=\sup_{G\in\mathcal{B}_B(0,n)} |\nu_x^n(G) -\nu_{x'}^n(G)|.
$$ If we consider stationary channels only (so that the definition of ADIMA can be made to require having only all $F \in \mathcal{B}_B(0,n)$ with the desired property, not with all $F \in \mathcal{B}_B(k,k+n)$ for all $k$), then we can rephrase the definition of the ADIMA channel by saying that a stationary channel is with ADIMA if, given $\epsilon > 0$, there are integers $m(\epsilon)$ and $a(\epsilon)$ such that for all n,
$$
v_n(\nu_x^n, \nu_{x'}^n) \leq \epsilon
$$ whenever $x_i = x'_i$, $-m \leq x \leq n+a$.

\subsection{$\bar{d}$-Distance and $\bar{d}$-Continuity}
Given two probability measures $\nu^n_x$ and $\nu^n_{x'}$ on $(B^n, \mathcal{B}_B(0,n))$, let $\mathcal{P}(\nu^n_x,\nu^n_{x'})$ denote the set of all joint probability measures $p^n$ on $(B^n \times B^n, \mathcal{B}_B(0,n)\times \mathcal{B}_B(0,n))$ satisfying the property that for all $G\in \mathcal{B}_B(0,n)$,
$$
p^n(G\times B^n) = \nu_x^n(G)
$$
and
$$
p^n(B^n\times G) = \nu_{x'}^n(G).
$$
In other words, $p^n$ has $\nu_x^n$ and $\nu_{x'}^n$ as marginals. Define the coordinate functions $U_i: B^n\times B^n \rightarrow B$ and $W_i: B^n\times B^n \rightarrow B$, $i, = 0, 1, \cdots, n-1$ by
$$
U_i(u^n,w^n) = u_i,
$$
$$
W_i(u^n,w^n) = w_i.
$$
Let $d_H(a,b) = 
  \begin{cases}
   1,\ a \neq b \\
   0,\ a = b   
  \end{cases}
$ be the Hamming distance and let $\delta_n$ to be the $n$th order normalized Hamming distance
$$
\delta_n(u^n,w^n) = \frac{1}{n}\sum^{n-1}_{i=0}d_H(u_i, w_i).
$$ The $n$th order $\bar{d}$-distance between $\nu_x^n$ and $\nu_{x'}^n$ is defined as
\begin{align*}
\bar{d}_n(\nu_x^n, \nu_{x'}^n) &= \inf_{p \in \mathcal{P}(\nu_x^n, \nu_{x'}^n)} E_p \delta_n(U^n,W^n) \\
                                                &= \inf_{p \in \mathcal{P}(\nu_x^n, \nu_{x'}^n)} \frac{1}{n} \sum_{i=0}^{n-1}p(U_i\neq W_i).
\end{align*}

A channel $\nu$ is $\bar{d}$-continuous if given $\epsilon >0$ there is an $n_0$ such that for $n \geq n_0$ we have $\bar{d}_n(\nu_x^n, \nu_{x'}^n) \leq \epsilon$ whenever $x_i = x'_i$, $i = 0,1, \cdots, n-1$. Equivalently, $\nu$ is  $\bar{d}$-continuous if
$$
\limsup_{n\rightarrow \infty} \sup_{a^n \in A^n} \sup_{x, x'\in c(a^n)} \bar{d}_n(\nu_x^n,\nu_{x'}^n) = 0.
$$

\subsection{Properties of $\bar{d}$-Continuous Channel}
Here we list several important properties of $\bar{d}$-continuous channel. We begin by establishing the relation between the ADIMA channel and $\bar{d}$-continuous channel.

\begin{theorem}[ADIMA Channel and $\bar{d}$-Continuity]
A channel $\nu$ is strongly $d$-continuous if given $\epsilon >0$ there are positive integers $m(\epsilon)$ and $a(\epsilon)$ such that for all $n$, $\bar{d}_n(\nu_x^n, \nu_{x'}^n) \leq \epsilon$ whenever $x_i = x'_i$, $-m \leq i \leq n-1+a$. An ADIMA channel is strongly $d$-continuous. A strongly $d$-continuous channel is $\bar{d}$-continuous.
\end{theorem}
\begin{IEEEproof}
See \cite{1056045}.
\end{IEEEproof}

Next, we summarize information-theoretic results. Recall the following notions:
\begin{align*}
C_{cb} \ \ &\text{Operational channel block coding capacity.}\\
C_{scb} \ \ &\text{Operational source/channel block  coding capacity.} \\
C^* \ \ &\text{Information quantile capacity.} \\
C_e\ \ &\text{Ergodic capacity.} \\
C_s \ \ &\text{Stationary capacity.} \\
C \ \ &\text{Block stationary capacity.}
\end{align*}
\begin{theorem}
a) For stationary channels we have
$$C = C_s = C_e.$$ Hence, for stationary channels, we can simply call $C = C_s = C_e$ the information rate capacity (or transmission rate capacity or Shannon capacity). \\
\\
b) If a channel is stationary and $\bar{d}$-continuous then 
$$
C \geq C^* = C_{scb} = C_{cb}.
$$
\\
c) Let $\nu$ be a discrete stationary $\bar{d}$-continuous channel. If $R < C^*$ and $\epsilon > 0$, then for sufficiently large $n_0$ there exist $(<2^{nR}>, n,\epsilon)$ block codes for all $n \geq n_0$. If $R > C^*$ there exists an $\epsilon_{\text{min}} >0$ such that there do not exist $(<2^{nR}>,n,\epsilon)$ block code for any $\epsilon < \epsilon_{\text{min}}$. \\
\\
d) Let $\nu$ be a discrete stationary $\bar{d}$-continuous channel and $\mu$ a discrete stationary ergodic source. If $H(\mu) < C^*$, then $\mu$ is admissible. If $H(\mu) > C^*$, then $\mu$ is not admissible. \\
\\
e) If, in addition, $\nu$ is also ergodic in b), c), and d), then $C=C^*$.
\end{theorem}
\begin{IEEEproof}
See \cite{1056045}.
\end{IEEEproof}

{\bf Theorem 4} lists the most important channel coding theorems for discrete stationary $\bar{d}$-continuous channels. For a discrete stationary $\bar{d}$-continuous channel {\bf Theorem 4.c)} says that if we transmit message at a rate $R<C^*$, then $R$ can be achieved by some block codes with arbitrarily small block error probability; conversely, if we transmit at a rate $R >C^*$, then the block error probability is always bounded away from zero. Similarly, {\bf Theorem 4.d)} says that, for a discrete stationary $\bar{d}$-continuous channel and any discrete stationary ergodic source $\mu$, if $H(\mu) < C^*$, then $\mu$ can be transmitted via block coding with arbitrarily small block error probability; conversely, if $H(\mu) > C^*$, then no such block code exists. The contents of {\bf Theorem 4.c)} and {\bf Theorem 4.d)} therefore collectively tell us that $C^*$ is the ultimate limitation of the system performance over the given channel, where the performance is measured in terms of the maximal code rate with vanishing block error probability. {\bf Theorem 4.e)} says that if, in addition, $\nu$ is also ergodic, then the Shannon capacity $C$ equals $C^*$.

The last property we need for $\bar{d}$-continuous channel is its preservation under the $\bar{d}$-limit \cite{1056074}:
\begin{theorem}
The $\bar{d}$-limit of a sequence of $\bar{d}$-continuous channels is $\bar{d}$-continuous.
\end{theorem}
\begin{IEEEproof}
See \cite{1056074}.
\end{IEEEproof}

\section{Channels in Diffusion-Based Molecular Communication}
This is the main section of our paper. In this section, we first state the general properties of diffusion-based molecular channel in Section IV.A. The abstract diffusion-based molecular channel model is given in Section IV.B and capacity theorems and channel coding theorems are proven in Section IV.C. Section IV.D discusses the implications and assumptions of our results.

\subsection{General Descriptions}
The theory of molecular communication studies the process of ``transmission and reception of information encoded in molecules'' \cite{akyildiznanonetworks}. In other words, in a molecular communication system, we utilize features of certain molecules as the information carrier, transmit the molecules through the communication media, and design a receiver which is capable of recognizing the features encoded over. The diffusion-based systems refer to those systems whose communication media are described by the ``diffusion processes''. In the following we introduce the essential components of diffusion-based molecular communication systems:  
\begin{figure}[h!]
\centering
\includegraphics[width=0.35\textwidth]{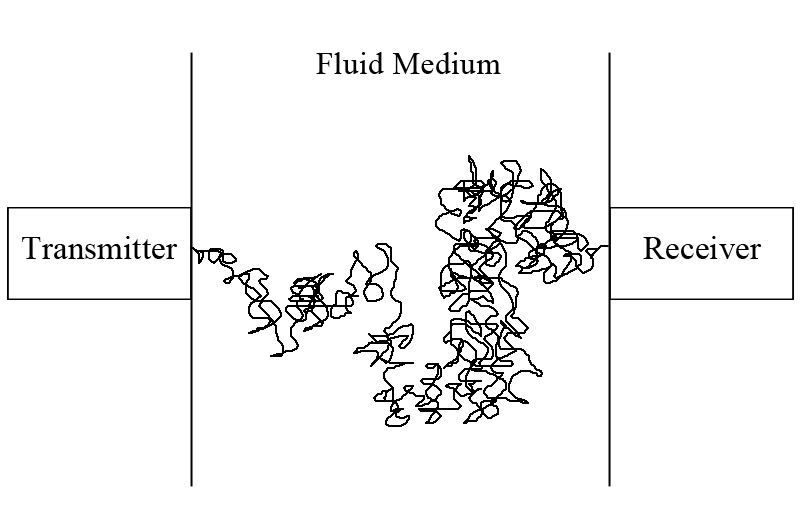}
\caption{Diffusion-Based Molecular Channel}
\label{V}
\end{figure}

\subsubsection{Transmitter}
In molecular communication, the transmitter has many choices of information carrier. For example, we can embed the information in the molecule concentration magnitudes \cite{6364540}, the inter-transmission times \cite{6364963}, the relative transmission times \cite{Eckford_2007}, the molecule types \cite{Eckford2007dis}, or hybrids of each other \cite{Hsieh2012}, \cite{6364963}. After the messages are encoded, the transmitter conveys the messages by simply releasing the encoded molecules into the communication media.\\

\subsubsection{Communication Media}
On contrast to the variety of information carrier, to the present days the communication media of molecular communication are concentrated on the fluid media almost exclusively. The fluid media represent the environment where the communication takes places. A typical environment is the blood, where the communication process occurs at human vessels. 

As is well-known, the motion of molecules in fluid media is described by the ``diffusion process''. The molecular communication systems whose communication media are described by the diffusion process are collectively called the ``free diffusion-based molecular systems'' or simply ``diffusion-based molecular systems''. For systems operating under the diffusion process, there is a special type of randomness which distinguishes those systems from the classical communication: the disorder of message arrivals. Consider an one-dimensional molecular communication system with transmitter and receiver separated by some distance (Fig. 1). For each released molecule diffusing independently in the fluid medium, the time it takes to be detected by the receiver is a random variable. If we assume that the receiver perfectly detects every molecule at its first arrival at the location of the receiver, then the detection times of the transmitted messages are the {\em first passage times} of the corresponding molecules, which are random variables depending only on the fluid medium, the corresponding transmitted molecules, and the distance from transmitter to receiver. Denote the first passage time of the $i$th message by $X_i$. It is possible to see that $X_i \leq X_j$ with positive probability even if $i>j$, i.e., it can happen that {\em the later transmitted message arrives earlier}. More generally, if we release a set of molecules in sequence, then it can happen that the receiver receives these molecules in {\em any} order.

The disorder of message arrivals is inevitable in diffusion-based molecular communication. Rather unfortunately, the classical tools dealing with disorders of message arrivals are targeting at the link-layer disorders (e.g. \cite{4797619}) and do not suit the case of diffusion-based disorder. Specifically, there are no existing mathematical tools describing the behavior of random ordering of arbitrary set of infinite random variables. To cope with it, we must develop useful channel model that describes the phenomena of the random ordering of first passage times, and by useful we mean the capability of applying the existing communication theory to the channel. The following is the strategy we take: we first find out the essential features of first passges times occurring in diffusion-based systems and define the notion of {\em permutation channel induced by the random ordering of first passage times}. Afterwards, we prove that the permutation channel is with ADIMA, and apply the channel coding theorems to such channel.

Despite the randomness described above, the first passage times in diffusion-based molecular systems do contain some regularities. The followings are the essential properties of first passage times in diffusion-based systems:
\begin{itemize}

\begin{figure}[h!]
\centering
\includegraphics[width=0.45\textwidth]{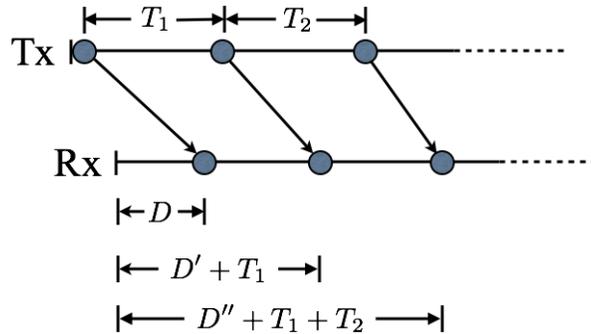}
\caption{First Passage Times}
\label{V}
\end{figure}
\item We require that the $X_{i+1}$ is a {\em delayed version} of $X_i$; that is, there are two independent random variables $D \geq 0$ and $D' \geq 0$ having the same distribution, and a random variable $T_i > 0$ independent of $D$ and $D'$, such that 
$$
X_i = D,
$$
$$
X_{i+1} = D' + T_i.
$$

The above condition actually says that we are releasing the encoded molecules {\em in sequence} (see Fig. 2). The intuitive meaning is that $T_i$ is the {\em inter-transmission} time of the $i$th and $(i+1)$th messages, and $D$ (or $D'$) is the first passage distribution of the $i$th message. If we release the $i$th message, wait for a time $T_i$, and release the $(i+1)$th message, then the time it takes for 
the $i$th and $(i+1)$th molecules to diffuse to the receiver is of the above form. Note that the above condition can be applied iteratively on $i$ to derive
$$
X_1 = D,
$$
$$
X_{i+1} = D' + \sum_{t=1}^i T_t,
$$ where $D$ and $D'$ are independent and have the same distribution, and all $T_i$'s are independent. Note also that the $X_i$'s are pairwise independent.

Let us give some examples. If the system is ``synchronous'' (which is the case in \cite{6364540}, \cite{6364963}, and \cite{Eckford_2007}) then $T_i = T$ for a real number $T>0$ and $X_{i+1} = X_1 + iT$. The work in \cite{6364963} also considers an asynchronous communication scheme for which $T_i$'s are i.i.d. random variables.\\

\item The second restriction we impose on the first passage times is a qualitative one: we shall require that the arrival order of each $X_i$ is {\em ``effected'' ``most exclusively''} by the $X_j$'s for those whose indices are close (or, equivalently, $|j-i|$'s are small). The precise meaning of ``effected'' and ``most exclusively'' are made clear by the next section, where we need the notion of the {\em permutation channel} to state them properly.

Our second restriction aims at capturing the nature of the diffusion processes occurring in molecular communication. The common feature of these diffusion processes is that if the release times of two molecules are $T_1$ and $T_2$, then the probability of the later released molecule advancing the first is very small provided $T_2 - T_1$ is large, {\em regardless of any feature of the transmitted molecules (e.g., weight, radius, ...)}. In other words, to infer the order of $X_j$ in $\{X_i\}_{i=1, 2, \cdots}$, it suffices to observe the nearby $X_{j-n}, X_{j-n+1}, \cdots, X_{j-1}$ and $X_{j+1}, X_{j+2}, \cdots, X_{j+n}$ for some large $n$. 

Let us give a concrete example to support the belief. Consider the diffusion process governed by the {\em Brownian motion with drift}, which is adopted in most diffusion-based  molecular communication systems (see, e.g., \cite{6364963}, \cite{Eckford_2007}, and \cite{Eckford2007dis}). The first passage time of a released molecule is given by \cite{Karlin1975}
$$
f_D(t) =   \begin{cases}
   \frac{x}{\sqrt{4\pi \nu}} \frac{1}{t^{\frac{3}{2}}}  \exp(-\frac{(vt-x)^2}{4\nu t}),\ t \geq 0 \\
   0,\ \ \ \ \ \ \ \ \ \ \ \  \ \ \ \ \  \ \ \ \ \ \ \ \ \ \  t < 0  
  \end{cases}\
$$ where $\nu$ is the diffusion coefficient, $x$ is the distance from the transmitter to the receiver, and $v$ is the drifting velocity. Suppose the inter-transmission times $T_i$'s are supported on $[t_i, \infty)$ and $\epsilon = \inf_i t_i >0$. Then, according to the first restriction, the probability of the $i$th released molecule advancing the first is
\begin{align*}
P(X_i \leq X_1) &= P(D' + \sum_{k=1}^i T_k \leq D) \\
                   &\leq P(D - D' \geq i\epsilon) \\
                   &\leq P(D \geq i\epsilon) \\
                   &\leq \int_{i\epsilon}^\infty f_D(u)du. 
\end{align*}
By discarding all the irrelevant constants we see that $f_D(u)$ behaves like $\frac{1}{u^{\frac{3}{2}}}  e^{-u}$ for large $u$. Therefore, for large $u$,
\begin{align*}
P(X_i < X_1) &\leq  \int_{i\epsilon}^\infty \frac{1}{u^{\frac{3}{2}}}  e^{-u} du \times \text{Constant}\\
                   &\leq (i\epsilon)^{-\frac{3}{2}} \times  \int_{i\epsilon}^\infty e^{-u} du \times \text{Constant}\\
                   &= (i\epsilon)^{-\frac{3}{2}} e^{-i\epsilon} \times \text{Constant}\\
                   &\rightarrow 0 \text{ as } i \rightarrow \infty.
\end{align*}
In fact, the above derivation shows that if $\epsilon \geq1$, then $e^i\times P(X_i < X_1) \rightarrow 0$ in $i$, which means that the probability of the $i$th release molecule advancing the first molecule converges to $0$ in $o(e^i)$ (that is, the probability decays super-exponentially). Similar reasonings apply to $P(X_i \leq X_j)$ with all $i, j$'s. In particular, if $\epsilon \geq 1$, then by union bound
\begin{align*}
P(X_j \leq X_i \text{ or } X_{j+1} \leq X_i \text{ or } \cdots) &\leq \sum_{s=j}^{\infty} P(X_s \leq X_i)\\
&\leq \sum_{s=j}^{\infty} e^{-s} \\
&\leq e\times e^{-j} \\ &\rightarrow 0 \text{ as } j\rightarrow \infty.
\end{align*}
That is, the probability of the event ``any molecule released after the $j$th message advancing the $i$th released molecule is vanishing in $j$''. This strong and universal property of diffusion processes is the one we shall capitalize upon when formulating the mathematical structure of diffusion-based molecular channels.\\
\end{itemize}

\subsubsection{Receiver}
In molecular communication, the receivers are designed to match the information carrier chosen by the transmitter. For example, if the information are carried on the molecule types, then the receiver is some nanomachines capable of recognizing these types. If the transmitter emits different concentration levels of a compound, then the receiver is usually operating on a chemical reaction involving the transmitted compound. Some examples for the first case are \cite{Hsieh2012}, \cite{6364963}, and \cite{Eckford2007dis}. For the later case, see, e.g., \cite{6364540} or \cite{Atakan2007}.

Just as the white noise may occur in classical communication, the receiver might contain some sources of detection noise. Again, in all cases of molecular communication, the detection noise of a message is governed by its nearby received messages ``most exclusively''. The mathematical formulation of a receiver is given in the next section.

\subsection{Channel Models and $\bar{d}$-Continuity}
For an integer $n$, denote the set of symmetric group of order $n$ by $S_n$; that is, $S_n$ is the set of all bijections of integers $\{1, 2, \cdots, n\}$ onto itself. An element $s \in S_n$ is called a {\em permutation} and can be represented by the permutation matrix:

\[
s = \left( \begin{array}[c]{ccccc} 1 & 2 & 3 & \cdots & n \\ 
                                            b_1 & b_2 & b_3 & \cdots & b_n
                \end{array} \right) \]  where $\{b_1, b_2, \cdots, b_n\}$ exhausts $\{1, 2, \cdots, n\}$, indicating the position after the effect of $s$; that is, $b_i = s(i),\ i = 1, 2, \cdots, n$. We also consider the {\em infinite permutation} defined as any bijection of the set of all integers $I = \{\cdots, -2, -1, 0, 1, 2\cdots\}$ onto itself. The set of all infinite permutations are denoted by $S_I$. Similar to $S_n$, any $s \in S_I$ has its permutation matrix representation:
\[
s = \left( \begin{array}[c]{ccccc} \cdots & -1 & 0 & 1 & \cdots \\ 
                                            \cdots & b_{-1} & b_0 & b_1 & \cdots
                \end{array} \right) \] for some doubly infinite sequence $\{b_i\}_{i\in I}$ as the position indicators.

Let $\gamma$ be any probability measure on $S_I$ (with Borel field being all subsets of $S_I$ in discrete topology). The effect of \[
s = \left( \begin{array}[c]{ccccc} \cdots & -1 & 0 & 1 & \cdots \\ 
                                            \cdots & b_{-1} & b_0 & b_1 & \cdots
                \end{array} \right) \] on an input sequence $x = (\cdots, x_{-1}, x_{0}, x_1, \cdots)$ is viewed as the permutation of index:
\begin{align}
s(x) = (x'), \ \ \ x'_i = x_{s^{-1}(i)} \text{ or equivalently } x'_{b_i} = x_i, \ \ i \in I,
\end{align} and $\gamma(s)$ is, of course, the probability of such index permutation.

Given a finite set of random variables $X_1, X_2, \cdots, X_n$, let $X_{(i)}$ denote the $i$th largest order statistic of $X_1, X_2, \cdots, X_n$; that is, $X_{(1)} \geq X_{(2)} \geq \cdots \geq X_{(n)}$. The order statistics of $X_{i}$'s induce a probability measure $\gamma$ on $S_n$ by considering the probability $\gamma(s)$ of \[
s = \left( \begin{array}[c]{ccccc} 1 & 2 & 3 & \cdots & n \\ 
                                            b_1 & b_2 & b_3 & \cdots & b_n
                \end{array} \right) \] as $P(X_1 = X_{(b_1)}, X_2=X_{(b_2)}, \cdots, X_n = X_{(b_n)})$. We shall say that $\gamma$ is induced by the ordering of $X_i$'s. For the case of $S_I$ the ordering of an infinite sequence of random variables is ambiguous. To this end, we need another way of identifying the bijections in $S_I$. For a finite subindex set $\{k_i\}_{i=1}^n$ (i.e., $k_i$'s are integers and $k_1 < k_2 < \cdots < k_n$) and a $s\in S_I$,\[ s = \left( \begin{array}[c]{ccccc} \cdots & -1 & 0 & 1 & \cdots \\ 
                                            \cdots & b_{-1} & b_0 & b_1 & \cdots
                \end{array} \right), \] define the {\em local $\{k_i\}_{i=1}^n$ version} of $s$ as 
\[
\left( \begin{array}[c]{ccccc} k_1 & k_2 & k_3 & \cdots & k_n \\ 
                                            b'_1 & b'_2 & b'_3 & \cdots & b'_n
            \end{array} \right) \] where $b'_i$ is the $i$th least integer of $b_{k_1}, b_{k_2}, \cdots, b_{k_n}$. For example, the local $\{1, 3, 5\}$ version of 
\[
\left( \begin{array}[c]{ccccc} 1 & 2 & 3 & 4 & 5 \\ 
                                            5 & 4 & 3 & 1 & 2
            \end{array} \right) \]
is 
\[
\left( \begin{array}[c]{ccc} 1 & 3 & 5 \\ 
                                         3 & 2 & 1 
            \end{array} \right) \]
since $5 > 3 > 2$. Obviously, for $s, s' \in S_I$, $s = s'$ if and only if every local version of $s$ and $s'$ coincides. Therefore, if we view the local $\{k_i\}_{i=1}^n$ version as the permutation matrix induced by the order statistics on the random variables $X_{k_1}, X_{k_2}, \cdots, X_{k_n}$, then these partial orderings of the infinite sequence $\{\cdots, X_{-1}, X_0, X_1, \cdots\}$ uniquely determine an infinite permutation $s \in S_I$. However, the explicit close-form solution of the distribution $\gamma$ on $S_I$ induced by $\{\cdots, X_{-1}, X_0, X_1, \cdots\}$ is a difficult and long standing problem for which all existing solutions require strong assumptions (see, e.g., the case for i.i.d. and exchangable random variables in \cite{csp}). Fortunately, we can take a different route to consider the {\em qualitative} properties of $\gamma$ and the qualitative features are enough for our application, as illustrated below.

We shall say that a probability measure $\gamma$ on $S_I$ is induced by the orderings of $\{\cdots, X_{-1}, X_0, X_1, \cdots\}$ if $\gamma$ possesses the following property:
for any finite subindex $\{k_i\}_{i=1}^n$ and any set of integers $b'_1, b'_2, \cdots, b'_n$ exhausting $\{1, 2, \cdots, n\}$, we have 
\begin{align}
P(X_{k_1} = X_{(b'_1)}, \cdots, X_{k_n} = X_{(b'_n)}) = \sum\gamma(s)
\end{align} where the summation is taken over all $s\in S_I$ with local $\{k_i\}_{i=1}^n$ version
\[
\left( \begin{array}[c]{ccccc} k_1 & k_2 & k_3 & \cdots & k_n \\ 
                                            b'_1 & b'_2 & b'_3 & \cdots & b'_n
            \end{array} \right). \]
In other words, for any finite sub-collection $\{X_{k_1}, \cdots, X_{k_n}\}$ of $\{\cdots, X_{-1}, X_0, X_1, \cdots\}$, we demand that the probability of the event $\{X_{k_1} = X_{(b'_1)}, \cdots, X_{k_n} = X_{(b'_n)}\}$ equals the sum of probability of the infinite permutations for which their local $\{k_i\}_{i=1}^n$ versions are the permutation matrix generated by the ordering of $\{X_{k_1}, \cdots, X_{k_n}\}$.

For difficulty illustrated before, the explicit construction for $\gamma$ satisfying the condition (2) is beyond the scope of this paper. In particular, the uniqueness of $\gamma$ is not of our concern: any $\gamma$ with property (2) is a good model for the channel under our consideration. If $\gamma$ is not unique we can simply choose any one as the representative.

We are now ready to define the notion of a {\em permutation channel}. Let $x = (\cdots, x_{-1}, x_0, x_1, \cdots)$ be any input sequence of a diffusion-based molecular communication system. As discussed in Section IV.A.2, $x$ corresponds a set of random variables $\{X_i\}_{i\in I}$ where $X_i$ is the first passage time of the message $x_i$. The ordering of these first passage times then induces a probability measure $\gamma_x$ on $S_I$. The {\em permutation channel $\{\nu_x\}_x$} on a same input and output alphabet, say $A$, is defined by
$$
\nu_x(y) = \gamma_x(s) , \ \ \ y = s(x), \ x, y \in A_{-\infty}^{\infty},
$$ where $s \in S_I$ and $y = s(x)$ means (1). With slight abuse of notation we also say that $\{\gamma_x\}_x$ or $\gamma$ is a permutation channel. \\

With the help of the permutation channel, we now give the following characterizations of a new class of channel, which is the mathematical channel model capturing all essential features of diffusion-based molecular channels considered in Section IV.A.

\begin{figure}[h!]
\centering
\includegraphics[width=0.45\textwidth]{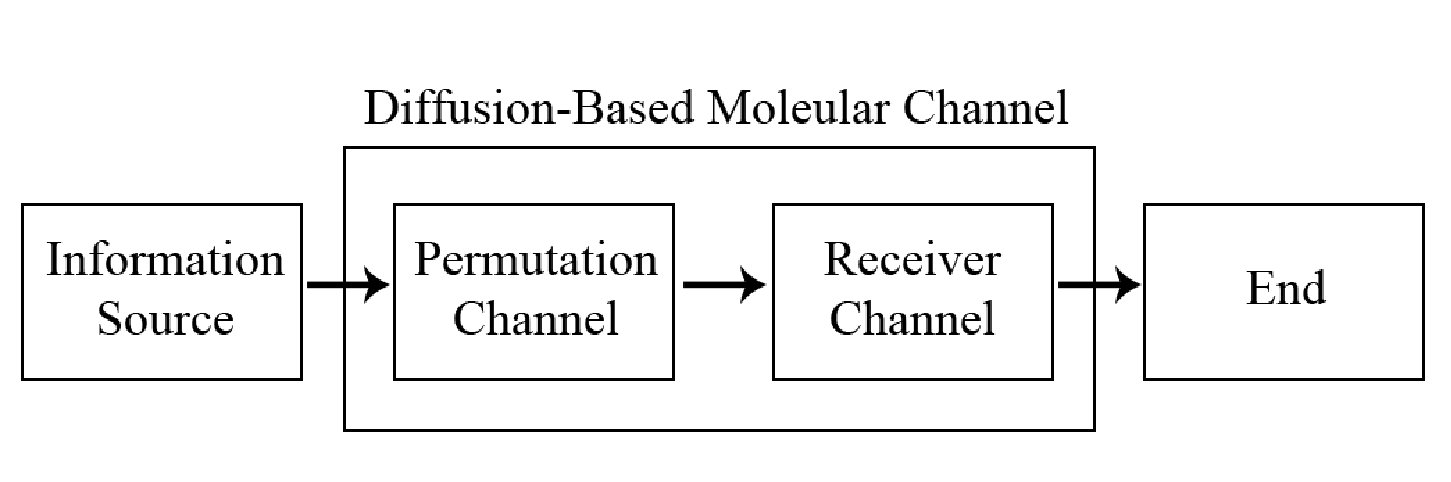}
\caption{Diffusion-Based Molecular Channel}
\label{V}
\end{figure}

\begin{definition}[The Diffusion-Based Molecular Channel]
A diffusion-based molecular channel $[A, \nu, B]$ is a cascade of a permutation channel $\gamma$ on $A$ and a stationary ergodic ADIMA channel $[A, \eta, B]$, where $\gamma$ is induced by the first passage times of input messages. $[A, \eta, B]$ is called the receiver channel. The permutation channel has the following property: Let $x$ and $x'$ be two input sequences and induce $\gamma_x$ and $\gamma_{x'}$ on $S_I$, respectively. For $\epsilon > 0$, there exists integers $m$ and $m'$ with $m' \leq m$ such that for all integer $n$, if $x_i = x'_i$ for $-m\leq i \leq n+m$, then the following holds:
\begin{description}
\item[(i)] For any $s \in S_I$ with $\sup_{i=0, 1, \cdots, n} s^{-1}(i) \leq m'+n$ and $\inf_{i=0, 1, \cdots, n} s^{-1}(i) \geq -m'$, we have 
$$
|\gamma_x(s) - \gamma_{x'}(s) |< \epsilon.
$$
\item[(ii)] The ``outlier'' probability is small:
\begin{align*}
P(\sup_{i=0, 1, \cdots, n} &s^{-1}(i) > m'+n \text{ or } \\ &\inf_{i=0, 1, \cdots, n} s^{-1}(i) < -m')< \epsilon
\end{align*} in both $\gamma_x$ and $\gamma_{x'}$ measure. In other words, 
$$
\sum_s \gamma_x(s) < \epsilon, \ \ \ \ \sum_s \gamma_{x'}(s) < \epsilon
$$ where the summation is taken over all $s$ with the property in the probability bracket.
\end{description}
\end{definition}

Let us give the engineering interpretation to all the assumptions we made. In section IV.A we have illustrated that the messages arrive at the receiver in disorder for diffusion-based molecular communication systems. Since such disorder comes from the random ordering of the first passage times, it does contain some implicit regularities. Suppose we aim at an interval $(y_0, y_1, \cdots, y_n)$ at the output of the permutation channel. Then condition (i) says that, since $x_i = x'_i$ for an interval $(-m, -m+1, \cdots, m+n)$ much larger than $(0, 1, \cdots, n)$, their corresponding first passage times are equally distributed within $(-m, -m+1, \cdots, m+n)$ and consequently the output distributions of $(y_0, y_1, \cdots, y_n)$ given $x$ and $x'$ should be close. Moreover, since the probability of $P(X_j \leq X_i)$ vanishes in $j - i$ in all diffusion processes in molecular communication, any model for a diffusion-based molecular channel must be able to capture the fact that the ``crossing probability $P(X_j \leq X_i)$'' is small in $j-i$, which is the content of condition (ii).

The total effect at the receiver is simply modeled as an ADIMA channel. The notion of ADIMA already characterizes fairly well a receiver whose detection of a message is ``mostly'' effected by its nearby received messages.

\subsection{Channel Capacity}
In this section we shall prove the following fundamental theorem of diffusion-based molecular communication:
\begin{theorem}[The Fundamental Theorem of Diffusion-Based Molecular Communication]
Let $A$ be a finite alphabet. For a diffusion-based molecular channel $[A, \nu, B]$ with a stationary ergodic source $[A, \mu]$,

a) $C = C_e = C_s = C^* = C_{cb} = C_{scb}.$\\

b) If $R < C$ and $\epsilon > 0$, then for sufficiently large $n_0$ there exist $(<2^{nR}>, n,\epsilon)$ block codes for all $n \geq n_0$. If $R > C$ there exists an $\epsilon_{\text{min}} >0$ such that there do not exist $(<2^{nR}>,n,\epsilon)$ block codes for any $\epsilon < \epsilon_{\text{min}}$. \\

c) If $H(\mu) < C$, then $\mu$ is admissible. If $H(\mu) > C$, then $\mu$ is not admissible.  \\
\end{theorem}

We divide our proof into the following steps:
\begin{description}
\item[(i)] The permutation channel in a diffusion-based molecular channel is stationary.
\item[(ii)] The permutation channel in a diffusion-based molecular channel is with ADIMA.
\item[(iii)] The cascade of ADIMA channels is an ADIMA channel.
\item[(iv)] By {\bf Theorem 3)}, an ADIMA channel is $\bar{d}$-continuous, and therefore  {\bf Theorem 4.a)$\sim$d)} apply.
\item[(v)] The diffusion-based molecular channel is strongly mixing (and therefore by  {\bf Theorem 2} it is ergodic).
\item[(vi)] The cascade of two ergodic channels is ergodic. By  {\bf Theorem 4.e)} the proof is complete.
\end{description}

\begin{IEEEproof}
(i): This is an immediate result of the fact that $Ty = s(Tx)$ if and only if $y = s(x)$, where $T$ is the time-shift operator.

(ii): Since by (i) the channel is stationary, it suffices to prove that for any $\epsilon >0$ and all integer $n$, there exists an integer $m=m(\epsilon)$ such that if $x_i = x'_i$ for $-m\leq i \leq n+m$, then $|\gamma_x^n(F) - \gamma_{x'}^n(F)| < \epsilon$ for any $F \in \mathcal{B}_A(0,n)$. Let $\epsilon > 0$ and $F$ be given. Let $m$ be a large integer determined later and $x_i = x'_i$ for $-m\leq i \leq n+m$. For each integer $m'$, divide the elements of $y_0^n \in F$ into three classes $F^{m'}_1$, $F'^{m'}_1$, and $F^{m'}_2$ where
\begin{align*}
F^{m'}_1 = &\{y_0^n\ |\ \exists s\in S_I \text{ with } \sup_{i=0, 1, \cdots, n} s^{-1}(i) \leq m'+n  \\ 
 &\text{ and }  \inf_{i=0, 1, \cdots, n} s^{-1}(i) \geq -m' \text{ such that } (s(x))_0^n = y_0^n\}, 
\end{align*}
\begin{align*}
F'^{m'}_1 = &\{y_0^n\ |\ \exists s\in S_I \text{ with } \sup_{i=0, 1, \cdots, n} s^{-1}(i) \leq m'+n  \\ 
 &\text{ and }  \inf_{i=0, 1, \cdots, n} s^{-1}(i) \geq -m' \text{ such that } (s(x'))_0^n = y_0^n\}, 
\end{align*}
and $F^{m'}_2$ is the complement of $F^{m'}_1 \cup F'^{m'}_1$. If $m' < m$, then $x_i = x'_i$ for $m' \leq i \leq m'+n$ and therefore $F^{m'}_1$ actually equals $F'^{m'}_1$. Now, by definition of a diffusion-based molecular channel, there is an $m$ such that if $y_0^n \in F^{m'}_1$, then
$$
|\gamma_x^n(y_0^n) - \gamma_{x'}^n(y_0^n)| <\epsilon
$$ since there exists a $s \in S_I$ with $\sup_{i=0, 1, \cdots, n} s^{-1}(i) \leq m'+n \text{ and }  \inf_{i=0, 1, \cdots, n} s(i) \geq -m'$ such that $(s(x))_0^n = (s(x'))_0^n$. On the other hand, if $y_0^n \in F^{m'}_2$, then either $\sup_{i=0, 1, \cdots, n} s^{-1}(i) > m'+n \text{ or } \inf_{i=0, 1, \cdots, n} s^{-1}(i) < -m'$. By definition of the diffusion-based molecular channel,
$$
|\gamma_x^n(y_0^n) - \gamma_{x'}^n(y_0^n)| \leq |\gamma_x^n(y_0^n)| + |\gamma_{x'}^n(y_0^n)| \leq 2\epsilon.
$$ Since any set $F \in \mathcal{B}_A(0,n)$ is a finite union of $y_0^n$, the proof is complete.

(iii): We shall state the result formally and prove it in appendix A.

(iv): It is a fact.

(v): Let $C_1 = C(y_0^n)$ and $C_2 = C(y_p^q)$ be two cylinder sets. The time indices $0, n, p,$ and $q$ are chosen for convenience and can be replaced by any other time. Without loss of generality suppose that $n-0 \geq q-p$. Since $T^k C(y_p^q) = C(T^k y_p^q) = C(y_{p+k}^{q+k})$,
\begin{align*}
P(C_1 \cap T^kC_2|x) &= P(C(y_0^n) \cap C(y_{p+k}^{q+k})|x) \\
&=\gamma_x(\{s\ |\ s(x)\in\ C(y_0^n) \cap C(y_{p+k}^{q+k})\}) \\
&=\gamma_x(C(y_0^n) \cap C(y_{p+k}^{q+k}) | s\in S^m_1)P(s\in S^m_1) \\ & \ \ + \gamma_x(C(y_0^n) \cap C(y_{p+k}^{q+k})|s \in S^m_2)P(s \in S^m_2)
\end{align*} where $C(y_0^n) \cap C(y_{p+k}^{q+k})$ means (of course) $\{s\ |\ s(x)\in\ C(y_0^n) \cap C(y_{p+k}^{q+k})\}$, and 
$$
S^m_1 = \{s \ |\ \sup_{i=0, 1, \cdots, n} s(i) \leq m+n \text{ and } \inf_{i=0, 1, \cdots, n} s(i) \geq -m\},
$$
$$
S^m_2 = (S^m_1)^c.
$$ By the definition of a diffusion-based molecular communication channel, there is a large $m$ such that $P(S^m_2) < \epsilon$ and therefore
$$
P(S^m_1) = 1- P(S^m_2) > 1- \epsilon.
$$ Now, choose $k>2m+n+p$ so that the intervals $(-m, -m+1, \cdots, m+n)$ and $(-m+p+k, -m+p+k+1, \cdots, m+q+k)$  are disjoint. Since 
\begin{align*}
s(x) \in C(y_0^n) \Leftrightarrow (s(x))_0^n = y_0^n,
\end{align*} for $s\in S^m_1$ this implies that $s(x) \in C(y_0^n)$ if and only if $y_0 = x_{i_0}, y_1 = x_{i_1}, \cdots, y_n = x_{i_n}$, $-m\leq i_0, i_1, \cdots, i_n \leq m+n$. By (2), the measure of the union of such $s$' is the probability of the event $\{X_0 = X_{i'_0}, X_1 = X_{i'_1}, \cdots, X_n = X_{i'_n}\}$ where the local $\{0, 1, \cdots, n\}$ version of $s$ is 
\[
\left( \begin{array}[c]{ccccc} 0 & 1 & 2 & \cdots & n \\ 
                                            i'_0 & i'_1 & i'_2 & \cdots & i'_n
            \end{array} \right). \] In other words,
$$
\gamma_x(C(y_0^n)\ |\ S^m_1) = P(\cup_{i'_0, \cdots, i'_n} \{X_0 = X_{(i'_0)}, \cdots, X_n = X_{(i'_n)}\})
$$ where $0\leq i'_0, \cdots, i'_n \leq n$ and $\{i'_0, \cdots, i'_n\}$ exhausts $\{0, 1, \cdots, n\}$.
Likewise, since the channel is stationary, for $s\in S^m_1$ we have $s(x) \in C(y_{p+k}^{q+k})$ if and only if $y_{p+k} = x_{i_{p+k}}, y_{p+k+1} = x_{i_{p+k+1}}, \cdots, y_{q+k} = x_{i_{q+k}}$, $-m+p+k\leq i_{p+k}, \cdots, i_{q+k} \leq m+k+q$, and
\begin{align*}
\gamma_x(C&(y_{p+k}^{q+k})\ |\ S^m_1) = \\ &P(\cup_{i'_0, \cdots, i'_{q-p}} \{X_{p+k} = X_{(i'_0)}, \cdots, X_{q+k} = X_{(i'_{q-p})}\}).
\end{align*}
Since the index sets of $\{X_i\}_{i=0}^n$ and $\{X_j\}_{j=p+k}^{q+k}$ are disjoint and all $X_i$'s are pairwise independent, we have, for large $k$,
\begin{align*}
& \gamma_x(C(y_0^n) \cap C(y_{p+k}^{q+k})|S^m_1)\\ & =P((\cup_{i'_0, \cdots, i'_n} \{X_0 = X_{(i'_0)}, \cdots, X_n = X_{(i'_n)}\})\cap \\ &\ \ \ \ \ \ (\cup_{i'_0, \cdots, i'_{q-p}} \{X_{p+k} = X_{(i'_0)}, \cdots, X_{q+k} = X_{(i'_{q-p})}\})) \\
&=P(\cup_{i'_0, \cdots, i'_n} \{X_0 = X_{(i'_0)}, \cdots, X_n = X_{(i'_n)}\})\times \\ &\ \ \ \ \ \ P(\cup_{i'_0, \cdots, i'_{q-p}} \{X_{p+k} = X_{(i'_0)}, \cdots, X_{q+k} = X_{(i'_{q-p})}\}) \\
&= \gamma_x(C(y_0^n)|S^m_1) \gamma_x(C(y_{p+k}^{q+k})|S^m_1),
\end{align*} and 
\begin{align*}
|P(C_1 \cap T^k C_2&|S^m_2)P(S^m_2) - \\ & P(C_1|S^m_2)P(S^m_2)P(T^kC_2|S^m_2)P(S^m_2)| \leq 2\epsilon.
\end{align*} Combining the above results shows that the permutation channel in a diffusion-based molecular channel is strongly mixing. By {\bf Theorem 2}, it is also ergodic.

(vi): By examining the definition of ergodicity of a channel, we see that, since the source together with the permutation channel can be viewed as another source to the receiver channel (which by definition is ergodic), the overall channel is ergodic. The above reasoning carries over to cascade of arbitrary number of ergodic channels.
\end{IEEEproof}

\subsection{Remarks on {\bf Theorem 6}}
The {\bf Theorem 6} provides the mathematical foundations for information theory in diffusion-based molecular communication. In this section, we discuss several issues concerning the assumptions of the {\bf Theorem 6} along with some important implications.
\begin{itemize}
\item Byproducts of {\bf Theorem 6}\\
The proof of {\bf Theorem 6} contains many useful byproducts. We list them formally as below:
\begin{theorem}[Stationarity of a Diffusion-Based Molecular Channel]
A diffusion-based molecular channel is stationary.
\end{theorem}
\begin{IEEEproof}
This is an easy consequence of the fact that the permutation channel and the receiver channel are both stationary, and the cascade of two stationary channels are stationary.
\end{IEEEproof}

\begin{theorem}[The Permutation Channel]
The permutation channel in a diffusion-based molecular channel is with ADIMA, strongly $d$-continuous, and $\bar{d}$-continuous.
\end{theorem}
\begin{theorem}[Ergodicity of the Diffusion-Based Molecular Channel]
A diffusion-based molecular channel is ergodic.
\end{theorem}

\begin{theorem}[$\bar{d}$-Continuity of the Diffusion-Based Molecular Channel]
A diffusion-based molecular channel is $\bar{d}$-continuous.\\
\end{theorem}

\item Why $\bar{d}$-Continuity? \\
In the proof of  {\bf Theorem 6} we have shown that the diffusion-based molecular channel is an ADIMA channel. Since ADIMA channel has its own coding theorems, it seems better to directly view the diffusion-based molecular channel as ADIMA channel and not bothering using the $\bar{d}$-continuity. This is, however, not appropriate due to the fact that the ADIMA channel is very ``vulnerable'' to modeling errors. This is best illustrated by the following example. Consider two i.i.d. binary sources with parameter $p$ and $q$. Recall that an ADIMA channel demands the variational distances between distributions under consideration to be close, which is {\em not} the case for these two i.i.d. sources when $p \neq q$ since the variational distance $v_n$ between the two i.i.d. sources tends to $1$ as $n\rightarrow \infty$, {\em no matter how small the error $|p - q|>0$ is}. The example tells us that if our model differs slightly from the reality, then the theorem might not carry over. Unfortunately, the diffusion process is itself a mathematical assumption and not the physical reality. In particular, for the most widely adopted diffusion model, the Brownian motion, various peculiar physical behaviors have been reported (e.g., almost-surely nowhere-differentiable path) and now it is commonly accepted that the Brownian motion is a good approximation of diffusion process but not the exact solution. There do exist realistic models for diffusion processes (e.g., the Ornstein-Uhlenbeck process), but they usually involve the ``stochastic differential equations'' which render the derivation of analytical results hard, if not impossible.

On contrast to the ADIMA channel, the $\bar{d}$-continuity is very ``robust'' to the modeling errors \cite{gra-neu-sh}. In particular, by {\bf Theorem 5} the $\bar{d}$-limit of a sequence of $\bar{d}$-continuous channels is $\bar{d}$-continuous, while the limit of a sequence of ADIMA channels is not necessarily an ADIMA channel. This tells us that suppose we can approximate the ``true'' channel by a sequence of $\bar{d}$-continuous channels, then the results we derived are still valid for the ``true'' channel. This is the reason we adopt the $\bar{d}$-continuity in this paper.\\

\item Continuous alphabets are possible. \\
In molecular communication, it can happen that the input and output alphabets are continuous (a classical example being the transmission of concentration waveforms over fluid media \cite{6364540}). We point out that there do exist works on the $\bar{d}$-continuous channel with continuous input and output alphabets \cite{Kadota:2006:CAM}. The extension of our results to continuous alphabets is therefore possible.\\

\item Why ergodic sources?\\
The notion of ergodicity comes from the statistical physics, where the introduce of ergodic process is used to describe a closed physical system. Since a large application area of molecular communication is to monitor the physical systems occurring in biological world, we are naturally lead to consider the ergodic sources. Besides, in most cases of communication a transmitter is itself a closed physical system, thus justifying the use of ergodicity. Last but not least, the ergodic sources have so many strong mathematical properties that we could barely derive anything without it. We just can't help but using it (and so do many great predecessors in information theory).\\

\item Why $C_{cb}$, $C^*$, and $C_{scb}$?\\
$C_{cb}$ clearly has its importance in coding theory. The importance of $C_{scb}$ and $C^*$ is not so obvious, as we illustrated below.

A major application of molecular communication is the monitoring of biological processes (e.g., the output signal of a human organ), which is usually assumed to be ergodic. For such scenarios, our communication problem at hand is ``Given an ergodic source $[A, \mu]$ and a diffusion-based molecular channel $[A, \nu, B]$, what is the best we can do?'' The question is answered in a very strong sense by the definition of $C_{scb}$ and {\bf Theorem 6}: a good source/channel code for $[A, \mu]$ exists if $H(\mu) < C_{scb}$; if $H(\mu) > C_{scb}$, no good source/channel code exist.

The reason for taking $C^*$ into consideration is of practical concern. Suppose that we try to really find out the channel capacity of the permutation channel $\gamma$. To this end, the direct calculation of $H(\mu)$, $H(\overline{\mu\gamma})$, and $H(\mu\gamma)$ is impractical, since the calculation of $H(\overline{\mu\gamma})$ and $H(\mu\gamma)$ involves all combinatorial terms for each $N$ (i.e., for each $N$ and a realization $(x_1, x_2, \cdots, x_N)$ and $(y_1, y_2, \cdots, y_N)$, we must consider all $s \in S_I$ for which $s(x) = y$) and the passing of limit $N\rightarrow\infty$. On contrast, since $C^*$ is defined through the {\em sample} mutual information $i_n(x^n, y^n)$, it suffices to choose a large $N$, calculate the expectation of $i_N(x^N, y^N)$, and the convergence of $i_n(x^n, y^n)$ in $L_1$ and the Shannon-McMillan Theorem will guarantee the closeness of $i_N(x^N, y^N)$ and $I(\mu\gamma)$.
\end{itemize}

\section{Conclusions}
As a promising paradigm for nano-communication, the molecular communication has been developed over the past decade. In this paper, we consider a major subclass of molecular communication systems called the diffusion-based molecular systems. Solid mathematical foundations for information theory are laid down for diffusion-based molecular communication. In particular, we have created an abstract channel model capturing all the essential features of diffusion-based molecular systems, and for such channel, the capacity theorems and channel coding theorems are proven. Other useful notions concerning stationarity and ergodicity of diffusion-based molecular channel are also established.

\appendices
\section{Cascade of ADIMA Channels}
Although the generalization is possible, here we shall confine ourselves to a special case which suffices our purpose.
\begin{theorem}[Cascade of ADIMA Channels]
For channels with finite input and output alphabets, the cascade of ADIMA channels is an ADIMA channel.
\end{theorem}
\begin{IEEEproof}
It suffices to consider the cascade of two ADIMA channels. The case for cascade of arbitrary number follows from induction. Let $[A, \sigma, Q]$ and $[Q, \eta, B]$ be two ADIMA channels with finite input and output alphabets and let $[A, \nu, B]$ be their cascade. Fix $\epsilon >0$. Since $\eta$ is with ADIMA, for any $n$ there are integers $m'$ and $a'$ such that for any $y \in B_0^n$,
\begin{align} |\eta_z(y) - \eta_{z'}(y)| \leq \epsilon \end{align}
if $z_i = z'_i$ for $-m' \leq i \leq n+a'$. Write
\begin{align*}
\nu_x(y) &= \int_{Q_{-\infty}^{\infty}} \eta_u(y) d\sigma_x(u)\\
              &= \sum_{z\in Q_{-m'}^{a'+n}} \int_{C(z)}\eta_u(y) d\sigma_x(u).
\end{align*} By (3), we have 
$$
\sup_{u, u' \in C(z)} |\eta_u(y) - \eta_{u'}(y)| \leq \epsilon.
$$ Put $\eta_{C(z)}(y) = \inf_{u \in C(z)}\eta_u(y)$. Then we can further write
\begin{align*}
\sum_{z\in Q_{-m'}^{a'+n}}\sigma_x(C(z)) &\eta_{C(z)}(y) \leq \nu_x(y) \\ &\leq \sum_{z\in Q_{-m'}^{a'+n}}\sigma_x(C(z))(\eta_{C(z)}(y) +\epsilon).\end{align*} Therefore,
\begin{align*}
|\nu_x(y) - \nu_{x'}(y)| &\leq \sum_{z\in Q_{-m'}^{a'+n}}|\sigma_x(C(z)) - \sigma_{x'}(C(z))| \times \text{Constant} \\
&\leq |Q_{-m'}^{a'+n}|\times \text{Constant}\times \epsilon
\end{align*} if $x$ and $x'$ are chosen with integers $m$ and $a$ such that $x_i = x'_i$ for $m \leq i \leq n+a$ implies $|\sigma_x(C(z)) - \sigma_{x'}(C(z))|$ is small, where the condition can be achieved since $[A, \sigma, Q]$ is with ADIMA. The proof is completed by noting that all $F \in \mathcal{B}_B(0,n)$ is a finite union of $y \in B_0^n$.
\end{IEEEproof}

\section*{Acknowledgment}
The authors would like to thank Prof. Ian F. Akyildiz for his inspring advices during a meeting in Taiwan. The authors would also like to thank Prof. Kwang-Cheng Chen, Prof. Chia-Han Lee, Po-Jen Shih, and Yen-Chi Lee for their contribution to this paper.


\begin{thebibliography}{10}
\providecommand{\url}[1]{#1}
\csname url@samestyle\endcsname
\providecommand{\newblock}{\relax}
\providecommand{\bibinfo}[2]{#2}
\providecommand{\BIBentrySTDinterwordspacing}{\spaceskip=0pt\relax}
\providecommand{\BIBentryALTinterwordstretchfactor}{4}
\providecommand{\BIBentryALTinterwordspacing}{\spaceskip=\fontdimen2\font plus
\BIBentryALTinterwordstretchfactor\fontdimen3\font minus
  \fontdimen4\font\relax}
\providecommand{\BIBforeignlanguage}[2]{{%
\expandafter\ifx\csname l@#1\endcsname\relax
\typeout{** WARNING: IEEEtran.bst: No hyphenation pattern has been}%
\typeout{** loaded for the language `#1'. Using the pattern for}%
\typeout{** the default language instead.}%
\else
\language=\csname l@#1\endcsname
\fi
#2}}
\providecommand{\BIBdecl}{\relax}
\BIBdecl

\bibitem{akyildiznanonetworks}
I.~F. Akyildiz, J.~M. Jornet, and M.~Pierobon, ``{Nanonetworks: a new frontier
  in communications},'' \emph{Communications of the ACM}, vol.~54, no.~11, pp.
  84--89, 2011.

\bibitem{citeulike:358722}
D.~L. Nelson and M.~M. Cox, \emph{{Lehninger Principles of Biochemistry, Fourth
  Edition}}, fourth edition~ed.\hskip 1em plus 0.5em minus 0.4em\relax W. H.
  Freeman, Apr. 2004.

\bibitem{Bossert1963443}
W.~H. Bossert and E.~O. Wilson, ``The analysis of olfactory communication among
  animals,'' \emph{Journal of Theoretical Biology}, vol.~5, no.~3, pp. 443 --
  469, 1963.

\bibitem{ParcerisaGine:2009:MCO}
L.~Parcerisa~Gin{\'e} and I.~F. Akyildiz, ``Molecular communication options for
  long range nanonetworks,'' \emph{Comput. Netw.}, vol.~53, no.~16, pp.
  2753--2766, Nov. 2009.

\bibitem{Nakano05molecularcommunication}
T.~Nakano, T.~Suda, M.~Moore, R.~Egashira, A.~Enomoto, and K.~Arima,
  ``Molecular communication for nanomachines using intercellular calcium
  signaling,'' in \emph{IEEE NANO 2005}, 2005.

\bibitem{6364540}
L.-S. Meng, P.-C. Yeh, K.-C. Chen, and I.~Akyildiz, ``A diffusion-based binary
  digital communication system,'' in \emph{Communications (ICC), 2012 IEEE
  International Conference on}, june 2012, pp. 4985 --4989.

\bibitem{Hsieh2012}
Y.-P. Hsieh, Y.-C. Lee, P.-J. Shih, P.-C. Yeh, and K.-C. Chen, ``On the
  asynchronous information embedding for event-driven systems in molecular
  communications,'' \emph{Nano Communication Networks}, no.~0, pp.~--, 2012.

\bibitem{6364963}
Y.-P. Hsieh, P.-J. Shih, Y.-C. Lee, P.-C. Yeh, and K.-C. Chen, ``An
  asynchronous communication scheme for molecular communication,'' in
  \emph{Communications (ICC), 2012 IEEE International Conference on}, june
  2012, pp. 6177 --6182.

\bibitem{6034228}
A.~Einolghozati, M.~Sardari, A.~Beirami, and F.~Fekri, ``Capacity of discrete
  molecular diffusion channels,'' in \emph{Information Theory Proceedings
  (ISIT), 2011 IEEE International Symposium on}, 31 2011-aug. 5 2011, pp. 723
  --727.

\bibitem{5935214}
M.~Pierobon and I.~Akyildiz, ``Information capacity of diffusion-based
  molecular communication in nanonetworks,'' in \emph{INFOCOM, 2011 Proceedings
  IEEE}, april 2011, pp. 506 --510.

\bibitem{6305481}
------, ``Capacity of a diffusion-based molecular communication system with
  channel memory and molecular noise,'' \emph{Information Theory, IEEE
  Transactions on}, vol.~59, no.~2, pp. 942 --954, feb. 2013.

\bibitem{shannon1948}
C.~E. Shannon, ``A mathematical theory of communication,'' \emph{The Bell
  System Technical Journal}, vol.~27, pp. 379--423, 623--, july, october 1948.

\bibitem{McMi53a}
B.~Mcmillan, ``{The Basic Theorems of Information Theory},'' \emph{Ann. Math.
  Stat.}, vol.~24, pp. 196--219, 1953.

\bibitem{Feinstein1954}
A.~Feinstein, ``A new basic theorem of information theory,'' 1954.

\bibitem{khinchin1957mathematical}
A.~Khinchin, \emph{Mathematical foundations of information theory}, ser. Dover
  books on advanced mathematics.\hskip 1em plus 0.5em minus 0.4em\relax Dover
  Publications, 1957.

\bibitem{1054666}
E.~Pfaffelhuber, ``Channels with asymptotically decreasing memory and
  anticipation,'' \emph{Information Theory, IEEE Transactions on}, vol.~17,
  no.~4, pp. 379 -- 385, jul 1971.

\bibitem{1056045}
R.~Gray and D.~Ornstein, ``Block coding for discrete stationary -continuous
  noisy channels,'' \emph{Information Theory, IEEE Transactions on}, vol.~25,
  no.~3, pp. 292 -- 306, may 1979.

\bibitem{genecapa}
S.~Verdu and T.~S. Han, ``{A general formula for channel capacity},''
  \emph{Information Theory, IEEE Transactions on}, vol.~40, pp. 1147--1157,
  1994.

\bibitem{1056074}
D.~Neuhoff and P.~Shields, ``Channels with almost finite memory,''
  \emph{Information Theory, IEEE Transactions on}, vol.~25, no.~4, pp. 440 --
  447, jul 1979.

\bibitem{Eckford_2007}
A.~Eckford, ``Nanoscale communication with brownian motion,'' \emph{Information
  Sciences and Systems CISS 07 41st Annual Conference on}, p.~6, 2007.

\bibitem{Eckford2007dis}
------, ``Achievable information rates for molecular communication with
  distinct molecules,'' in \emph{Bio-Inspired Models of Network, Information
  and Computing Systems, 2007. Bionetics 2007. 2nd}, dec. 2007, pp. 313 --315.

\bibitem{4797619}
J.~Walsh and S.~Weber, ``Capacity region of the permutation channel,'' in
  \emph{Communication, Control, and Computing, 2008 46th Annual Allerton
  Conference on}, 2008, pp. 646--652.

\bibitem{Karlin1975}
S.~Karlin and H.~M. Taylor, \emph{{A First Course in Stochastic Processes,
  Second Edition}}, 2nd~ed.\hskip 1em plus 0.5em minus 0.4em\relax Academic
  Press, Apr. 1975.

\bibitem{Atakan2007}
B.~Atakan and O.~B. Akan, ``{An information theoretical approach for molecular
  communication},'' in \emph{2007 2nd Bio-Inspired Models of Network,
  Information and Computing Systems}.\hskip 1em plus 0.5em minus 0.4em\relax
  IEEE, 2007, pp. 33--40.

\bibitem{csp}
J.~Pitman, \emph{Combinatorial stochastic processes}, ser. Lecture Notes in
  Mathematics.\hskip 1em plus 0.5em minus 0.4em\relax Berlin: Springer-Verlag,
  2006, vol. 1875.

\bibitem{gra-neu-sh}
R.~M. Gray, D.~L. Neuhoff, and P.~C. Shields, ``A generalization of the
  ornstein's distance with application to information theory,'' \emph{Annals of
  Probability}, vol.~3, Apr. 1975.

\bibitem{Kadota:2006:CAM}
T.~Kadota, ``On the capacity of asymptotically memoryless continuous-time
  channels (corresp.),'' \emph{IEEE Trans. Inf. Theor.}, vol.~19, no.~4, pp.
  556--557, Sep. 2006.

\bibitem{Akyildiz2010}
I.~F. Akyildiz and J.~M. Jornet, ``The internet of nano-things,''
  \emph{Wireless Commun.}, vol.~17, pp. 58--63, 2010.

\bibitem{Pierobon2010}
M.~Pierobon and I.~F. Akyildiz, ``A physical end-to-end model for molecular
  communication in nanonetworks,'' \emph{IEEE J.Sel. A. Commun.}, vol.~28, pp.
  602--611, May 2010.

\bibitem{Atakan2008}
B.~Atakan and O.~B. Akan, ``On molecular multiple-access, broadcast, and relay
  channels in nanonetworks,'' in \emph{Proceedings of the 3rd International
  Conference on Bio-Inspired Models of Network, Information and Computing
  Sytems}, ser. BIONETICS '08.\hskip 1em plus 0.5em minus 0.4em\relax ICST
  (Institute for Computer Sciences, Social-Informatics and Telecommunications
  Engineering), 2008, pp. 16:1--16:8.

\bibitem{WilliamsRogers}
L.~C.~G. Rogers and D.~Williams, \emph{Diffusions, {M}arkov Processes, and
  Martingales. {V}ol. 1}.\hskip 1em plus 0.5em minus 0.4em\relax Cambridge:
  Cambridge University Press, 2000.

\bibitem{KadloorA09}
S.~Kadloor and R.~Adve, ``A framework to study the molecular communication
  system.'' in \emph{ICCCN}.\hskip 1em plus 0.5em minus 0.4em\relax IEEE, 2009,
  pp. 1--6.

\bibitem{Dudley2002}
R.~M. Dudley, \emph{{Real Analysis and Probability}}, 2nd~ed.\hskip 1em plus
  0.5em minus 0.4em\relax Cambridge University Press, Aug. 2002.

\bibitem{Atakan2008CCE}
B.~Atakan and O.~B. Akan, ``On channel capacity and error compensation in
  molecular communication,'' in \emph{Transactions on Computational Systems
  Biology}.\hskip 1em plus 0.5em minus 0.4em\relax Springer-Verlag, 2008, pp.
  59--80.

\bibitem{Yeung2006}
R.~W. Yeung, \emph{A First Course in Information Theory (Information
  Technology: Transmission, Processing and Storage)}.\hskip 1em plus 0.5em
  minus 0.4em\relax Secaucus, NJ, USA: Springer-Verlag New York, Inc., 2006.

\bibitem{Gray:2009:PRP:1795714}
R.~M. Gray, \emph{Probability, Random Processes, and Ergodic Properties},
  2nd~ed.\hskip 1em plus 0.5em minus 0.4em\relax Springer Publishing Company,
  Incorporated, 2009.

\bibitem{Gray:1990:EIT:90455}
------, \emph{Entropy and information theory}.\hskip 1em plus 0.5em minus
  0.4em\relax New York, NY, USA: Springer-Verlag New York, Inc., 1990.

\bibitem{6089591}
A.~Einolghozati, M.~Sardari, and F.~Fekri, ``Capacity of diffusion-based
  molecular communication with ligand receptors,'' in \emph{Information Theory
  Workshop (ITW), 2011 IEEE}, oct. 2011, pp. 85 --89.

\bibitem{Akyildiz:2008ACM}
I.~F. Akyildiz, F.~Brunetti, and C.~Bl\'{a}zquez, ``Nanonetworks: A new
  communication paradigm,'' \emph{Comput. Netw.}, vol.~52, no.~12, pp.
  2260--2279, Aug. 2008.

\bibitem{AkyildizEM}
I.~F. Akyildiz and J.~M. Jornet, ``{Electromagnetic wireless nanosensor
  networks},'' \emph{Nano Communication Networks}, vol.~1, no.~1, pp. 3--19,
  Mar. 2010.

\bibitem{Alfano2006}
G.~Alfano and D.~Miorandi, ``{On Information Transmission Among
  Nanomachines},'' in \emph{Nano-Networks, (Nano-Net '06), 1st International
  Conference on}, 2006.

\bibitem{Arifler11}
D.~Arifler, ``Capacity analysis of a diffusion-based short-range molecular
  nano-communication channel.'' \emph{Computer Networks}, vol.~55, no.~6, pp.
  1426--1434, 2011.

\bibitem{citeulike:9457719}
B.~Atakan and O.~B. Akan, ``{Deterministic capacity of information flow in
  molecular nanonetworks},'' \emph{Nano Communication Networks}, vol.~1, no.~1,
  pp. 31--42, Mar. 2010.

\bibitem{Costerton1999}
J.~W. Costerton, P.~S. Stewart, and E.~P. Greenberg, ``Bacterial biofilms: A
  common cause of persistent infections,'' \emph{Science}, vol. 284, no. 5418,
  pp. 1318--1322, 1999.

\bibitem{5875906}
M.~Pierobon and I.~Akyildiz, ``Noise analysis in ligand-binding reception for
  molecular communication in nanonetworks,'' \emph{Signal Processing, IEEE
  Transactions on}, vol.~59, no.~9, pp. 4168 --4182, sept. 2011.

\bibitem{Pierobon:2011DNA}
M.~Pierobon and I.~F. Akyildiz, ``Diffusion-based noise analysis for molecular
  communication in nanonetworks,'' \emph{Trans. Sig. Proc.}, vol.~59, no.~6,
  pp. 2532--2547, Jun. 2011.

\bibitem{Moore:2006:DMC}
M.~Moore, A.~Enomoto, T.~Nakano, R.~Egashira, T.~Suda, A.~Kayasuga, H.~Kojima,
  H.~Sakakibara, and K.~Oiwa, ``A design of a molecular communication system
  for nanomachines using molecular motors,'' in \emph{Proceedings of the 4th
  annual IEEE international conference on Pervasive Computing and
  Communications Workshops}, ser. PERCOMW '06.\hskip 1em plus 0.5em minus
  0.4em\relax Washington, DC, USA: IEEE Computer Society, 2006, pp. 554--.

\bibitem{Kuran201265}
M.~S. Kuran, H.~B. Yilmaz, T.~Tugcu, and I.~F. Akyildiz, ``Interference effects
  on modulation techniques in diffusion based nanonetworks,'' \emph{Nano
  Communication Networks}, vol.~3, no.~1, pp. 65 -- 73, 2012.

\bibitem{Freitas06pharmacytes}
R.~A. Freitas, ``Pharmacytes: An ideal vehicle for targeted drug delivery,''
  \emph{Journal of Nanoscience and Nanotechnology}, pp. 2769--2775, 2006.

\bibitem{cussler1997diffusion}
E.~Cussler, \emph{Diffusion: Mass Transfer in Fluid Systems}, ser. Cambridge
  Series in Chemical Engineering.\hskip 1em plus 0.5em minus 0.4em\relax
  Cambridge University Press, 1997.

\end{thebibliography}
\end{document}